\documentclass[review,11pt]{elsarticle}

\usepackage{lineno,hyperref,geometry}
\modulolinenumbers[5]

\journal{International Journal of Hydrogen Energy}

\makeatletter
\def\ps@pprintTitle{%
 \let\@oddhead\@empty
 \let\@evenhead\@empty
 \def\@oddfoot{}%
 \let\@evenfoot\@oddfoot}
\makeatother
\begin{document}

\begin{frontmatter}

\title{A new palladium alloy with near-ideal hydrogen storage performance}

\author[ikst]{Pritam Das}

\author[kist2]{Young-Su Lee}

\author[kist]{Seung-Cheol Lee \corref{correspondingauthor}}
\ead{leesc@kist.re.kr}

\cortext[correspondingauthor]{Corresponding author}

\author[ikst]{Satadeep Bhattacharjee \corref{correspondingauthor}}
\ead{s.bhattacharjee@ikst.res.in}

\address[ikst]{Indo-Korea Science and Technology Center (IKST), Jakkur, Bengaluru 560065, India}
\address[kist]{Electronic Materials Research Center, KIST, Seoul 136-791, South Korea}
\address[kist2]{Center for Energy Materials Research, Korea Institute of Science and Technology, Seoul 02792, Seoul, Republic of Korea.}

\begin{abstract}
Hydrogen-based fuels demand high-density storage that can operate under ambient temperatures. Pd and its alloys are the most investigated metal hydrides for hydrogen fuel cell applications. This study presented an alternative Pd alloy for hydrogen storage that can store and release hydrogen at room temperature. The surface of the most studied Pd (110) was modified with Au and Rh so that the hydrogen adsorption energy was 0.49 eV and the release temperature was 365 K. Both values are quite near to the optimum values for the adsorption energy and release temperature of a hydrogen fuel cell in real-world usage.
\end{abstract}


\end{frontmatter}


\section*{Introduction}

Given the urgent need to replace fossil fuels, hydrogen has emerged as an appealing energy carrier due to its high gravimetric energy density.  Hydrogen fuel cells have gained particular attention in transportation sectors including heavy vehicles. Hydrogen can be produced from naturally abundant and green sources \cite{Dutta2014}. One of the challenges in hydrogen fuel cell applications is hydrogen storage technology. There are mainly three categories for hydrogen storage technologies, namely: (1) physical storage in which hydrogen might be stored as compressed gas or a fluid in pure molecular form with no critical physical or chemical bonding to different materials; (2) adsorption in which hydrogen might be adsorbed onto or into a material, held by feeble physical van der Waals bonds; and (3) chemical storage in which hydrogen might be synthetically fortified (absorbed) \cite{Andersson2019}. Compressed hydrogen requires the use of expensive and potentially unsafe type IV carbon fiber-reinforced composite tanks to load hydrogen up to about 700 bar \cite{DURBIN2013}. Whereas, liquid hydrogen tanks are refrigerated at a temperature of $\sim$20 K to liquefy hydrogen at atmospheric pressure. In both cases, these approaches are complicated and costly, and their extreme operating conditions raise safety concerns \cite{Allendorf2018}. Among the existing approaches, solid-state storage of hydrogen has the potential to provide the highest gravimetric and volumetric hydrogen storage densities and may become the most suitable method for practical implementation of hydrogen storage. Common solid-state hydrogen storage materials are metal hydrides, complex hydrides, and chemical hydrides \cite{YU2017, SUN2018}.\\

Metal hydride is becoming a popular way to store hydrogen because it can hold more hydrogen at room temperature with less risk \cite{Sarac2022, Colbe2019}. Among different metals, Pd and its alloys gained popularity because of the high affinity between Pd and hydrogen \cite{Kansara2021}. Many research groups have studied the adsorption, absorption, and desorption of hydrogen on Pd and its alloys \cite{Namba2018, Padama2015, Padama2012, Sykes2005}. Depending on the dimensions, hydrogen storage capacity changes in the Pd nanoparticle \cite{Yamauchi2008}. Surface morphology and chemistry also play an important role in the hydrogen storage performance of Pd nanoparticles \cite{Xu2013}. A number of studies were carried out using a combination of Pd based alloys such as Au/Pd, Cu/Pd, Mg/Pd, graphene/Pd, Pt/Pd, and Cd/Pd \cite{EkborgTanner2021, Adams2009, Zhou2016, TAKEICHI2007, Yamauchi2009, Sonwane2006}. Most of the studies focused on hydrogen storing capacity and its improvement. Like, Sonwane et al. changes Au and Ag concentration in Pd alloys and found H has maximum solubility at $\sim20$\%\ Ag and $\sim12$\%\ Au concentration in Pd alloys \cite{Sonwane2006}. Hydrogen absorbed from the surface and stored in the second subsurface layer Ag/Pd and the Ag segregated to the topmost layer is energetically more favourable than the clean surface structure \cite{PADAMA2013}. Au/Pd alloys show higher hydrogen solubility than pure Pd. In the Au/Pd alloy, lowering the barrier for H atoms to get from the surface to the second subsurface speeds up the process of hydrogen absorption \cite{Namba2018}. Another technological challenge is effectively adsorbing and releasing hydrogen from the fuel cell. Hydrogen adsorption and release energy and temperature can be controlled by changing the structure and surface chemistry of the material. Very few studies are available on the adsorption and release mechanisms of hydrogen storage. Williamson et al. replaced silicon atoms with carbon in the Si\textsubscript{29} cluster, which reduces the hydrogen release temperature from 580 K to 380 K \cite{Williamson2004}. The hydrogen release temperature decreases due to the shorter C-C bond in Si\textsubscript{29} cluster, which reduces separation and increases repulsion between neighbouring silicon dihydrides, favouring reconstruction of the dimer structure and lowering the release temperature. To our knowledge, no studies on the hydrogen release mechanism on Pd alloy have been performed.\\

In the present study, an alternative Pd structure has been proposed for hydrogen storage in which hydrogen can be stored and released at ambient conditions. From our first-principles based approach we design the Pd structure that operates at atmospheric temperature and pressure from the most studied Pd (110) surface \cite{Namba2018}. In addition, the critical hydrogen chemical potential of Au/Pd and Rh/Pd was also examined. In the following section, we briefly discuss the computational details before discussing our results and conclusions.

\section*{Computational details}

All calculations have been performed using the density functional theory (DFT) as implemented in the VASP (Vienna Ab initio Software Package) code \cite{Kresse1996CMS, Kresse1996PRB}. For the core and valence electrons, PAW (projected augmented wave) potentials and plane-wave basis sets were used with the Perdew–Burke–Ernzerhof (PBE) functional to get the electronic energy \cite{Kresse1999, Bl1994}. Corrections were made to account for long-range interactions using the semi-empirical Grimme D2 dispersion method and for non-spherical contributions to the PAW potentials that were built into the code \cite{Grimme2006}. All energies converged within a cutoff of 500 eV. The conjugate gradient algorithm is used for structural optimization \cite{Perdew1996}. The convergence criteria for energy and force are 10\textsuperscript{-7} eV and -0.01 eV\AA\textsuperscript{-1}, respectively. In all calculations, spin polarisation was enabled, with the only exception of the isolated H\textsubscript{2} closed shell molecule.\\

The (110) surface of the Pd in the study was simulated by $2\times3\times6$ supercell slab models (Fig. 1 (a)). The clean surface is initially optimised before introducing hydrogen, gold, and rhodium atoms. The four topmost atomic layers are allowed to relax while keeping the bottom layers fixed in their bulk parameters. A vacuum of 15 \AA ~thickness along the z-direction is employed to avoid interactions between the neighbouring layers. For hydrogen absorption and desorption studies, half of the surface Pd atoms of Pd (110) were periodically replaced by Au and Rh atoms. Three different Pd (110) surfaces were formed. In the first two structures, half of the surface Pd atoms were replaced by Au and Rh atoms, and the structures were denoted as Au/Pd and Rh/Pd (Fig.1 (b)), respectively. In the third structure, three Pd atoms are replaced by two Au and one Rh atom, and the structure is denoted as Au+Rh/Pd (Fig.1 (c)).\\

\begin{figure}[h!]
	\centering
	\includegraphics[width=0.75\linewidth]{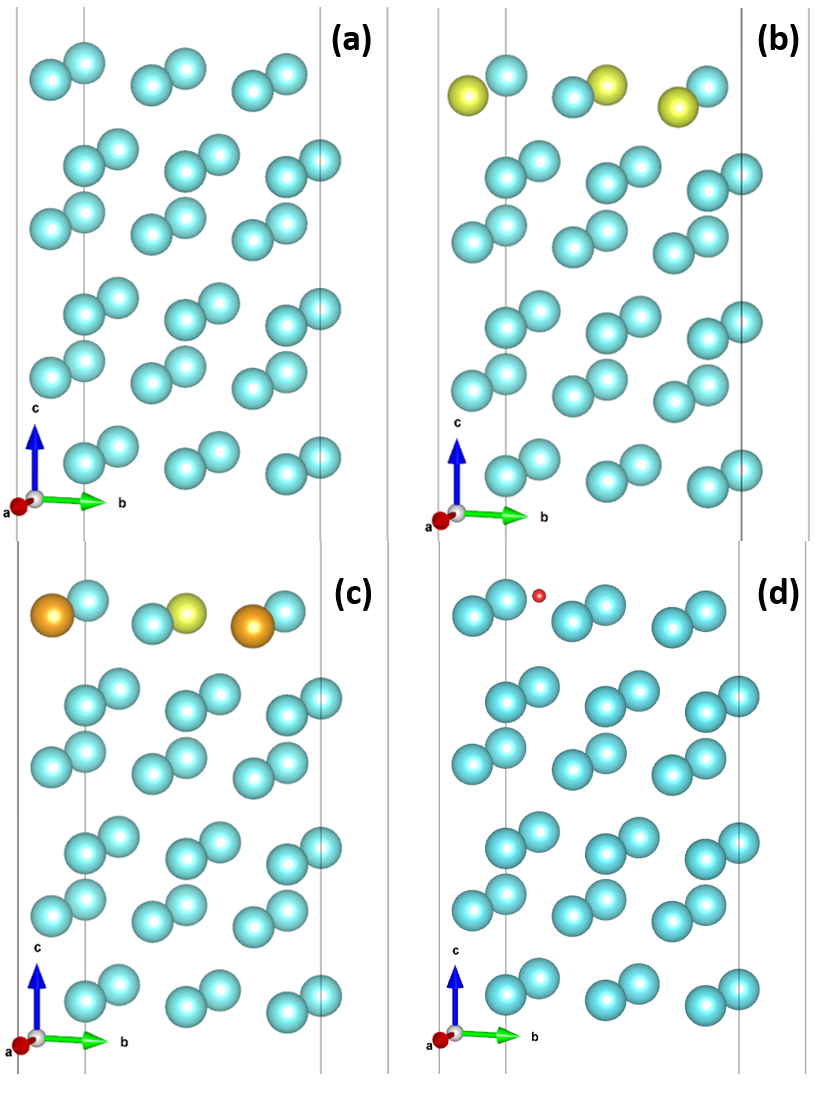}
	\caption{Optimized structure of (a) Pd (110) surface, (b) Rh/Pd surface, (c) Au+Rh/Pd surface, and (d) Pd (110) surface with hydrogen at octahedral site. Blue, yellow, green, and red balls correspond to Pd, Au, Rh, and H, respectively.}
	\label{fig:1}
\end{figure}

The absorption energy (E\textsubscript{abs}) of hydrogen on the different surfaces were calculated by: 

\begin{equation}
	E_{abs} =\frac{1}{n}\times\left(E_{slab+nH}-\left( E_{slab}+\frac{n}{2}E_{H_2}\right)\right)
\end{equation}

where E\textsubscript{slab+nH} is the energy of the hydrogenated slab, E\textsubscript{slab} and E\textsubscript{H\textsubscript{2}} are the energies of the pristine surface and isolated H\textsubscript{2}, respectively, and n is the number of hydrogen atoms involved in the adsorption.\\

The atomic charges for all structures were calculated by Bader charge analysis, as suggested by Henkelman and co-workers \cite{Henkelman2006, Tang2009}. The work functions ($\phi$) were calculated by subtracting the Fermi energy from the vacuum energy, extracted from the local potential by averaging the energy levels above the surface. In order to estimate the release temperature, the chemical potential of the hydrogen in the solid phase and in the gas phase needs to be evaluated. The chemical potential energy of hydrogen as a function of temperature (T) and pressure (p) was obtained using standard statistical thermodynamics, via the formula \cite{Khan2016}:

\begin{equation}
	\mu(T,p)=\frac{1}{2}\left(E_{el}+E_{ZPE}+H^0(T)-H^0(0)-TS^0(T)+k_BTln\left(\frac{p}{p^0}\right)\right)
\end{equation}

where E\textsubscript{el} and E\textsubscript{ZPE} are the electronic and zero point energy of the hydrogen molecule as derived from DFT, H\textsuperscript{0} and S\textsuperscript{0} is  enthalpy and entropy at standard pressure p\textsuperscript{0} = 1 bar. The values of enthalpy and entropy at standard pressure can be found in JANAF thermochemical tables \cite{Chase1982}. The critical hydrogen chemical potential, $\mu^C_H$, at which the different Pd structure release hydrogen from the surface is given by $\mu^C_H=1/n(E_{slab+nH}-E_{slab})$ \cite{Williamson2004}.

\section*{Results and discussion}
The hydrogen adsorption was studied considering only the surface octahedral site as it has been shown in the previous study that hydrogen in the subsurface octahedral and tetrahedral sites are less stable. To calculate the adsorption energies, we place the hydrogen atom  on such surface octahedral site for different surfaces as shown in the  Fig.\ref{fig:1} (d) Hydrogen atoms are exothermically adsorbed on all surfaces. But to effectively store hydrogen in a storage medium, the ideal hydrogen adsorption energy should be approximately -0.42 eV/atom \cite{Lochan2006}. It can be seen that for the clean Pd (110) surface and Rh doped Pd (110) surface (Rh/Pd) have much higher absorption energies, -0.56 and -0.63 eV, respectively. Whereas  adsorption energy of the Au/Pd and Au+Rh/Pd surfaces are very close to the ideal adsorption energy, at -0.40 and -0.49 eV, respectively. These results are shown in the table-\ref{table:1}. A Bader charge analysis (table-\ref{table:1}) has been performed on all the hydrogenated surfaces. The electron-donating nature of the metal is clearly observed on all surfaces. 

\begin{table}[h!]
	\caption{The adsorption energy and Bader charges on hydrogen (in atomic unit) on different Pd surface.}
	\centering
	\begin{tabular}{c c c} 
		\hline
		\textbf{Surface} & \textbf{E\textsubscript{ads}} & \textbf{e\textsuperscript{-}} \\
		\hline
		Pd & -0.56 & 1.13 \\
		Au/Pd & -0.40 & 1.09 \\
		Rh/Pd & -0.63 & 1.18 \\
		Au+Rh/Pd & -0.49 & 1.14 \\
		\hline
	\end{tabular}
	\label{table:1}
\end{table}

The hydrogen adsorption energy has a clear correlation with the d-band centers of the metal surfaces. The surfaces show an inverse correlation between adsorption energy and d-band centers(Fig. 2). This behaviour quite consistent with Newns-Anderson model or d-band center model of chemisorption on the metal surfaces~\cite{newns,d1,d2}. A similar trend is also observed between charges on hydrogen and d-band position. The surface with higher d-band center exhibits lower charge donation to hydrogen and smaller (less negative) adsorption energy.  \\

\begin{figure}[h!]
	\centering
	\includegraphics[width=0.75\linewidth]{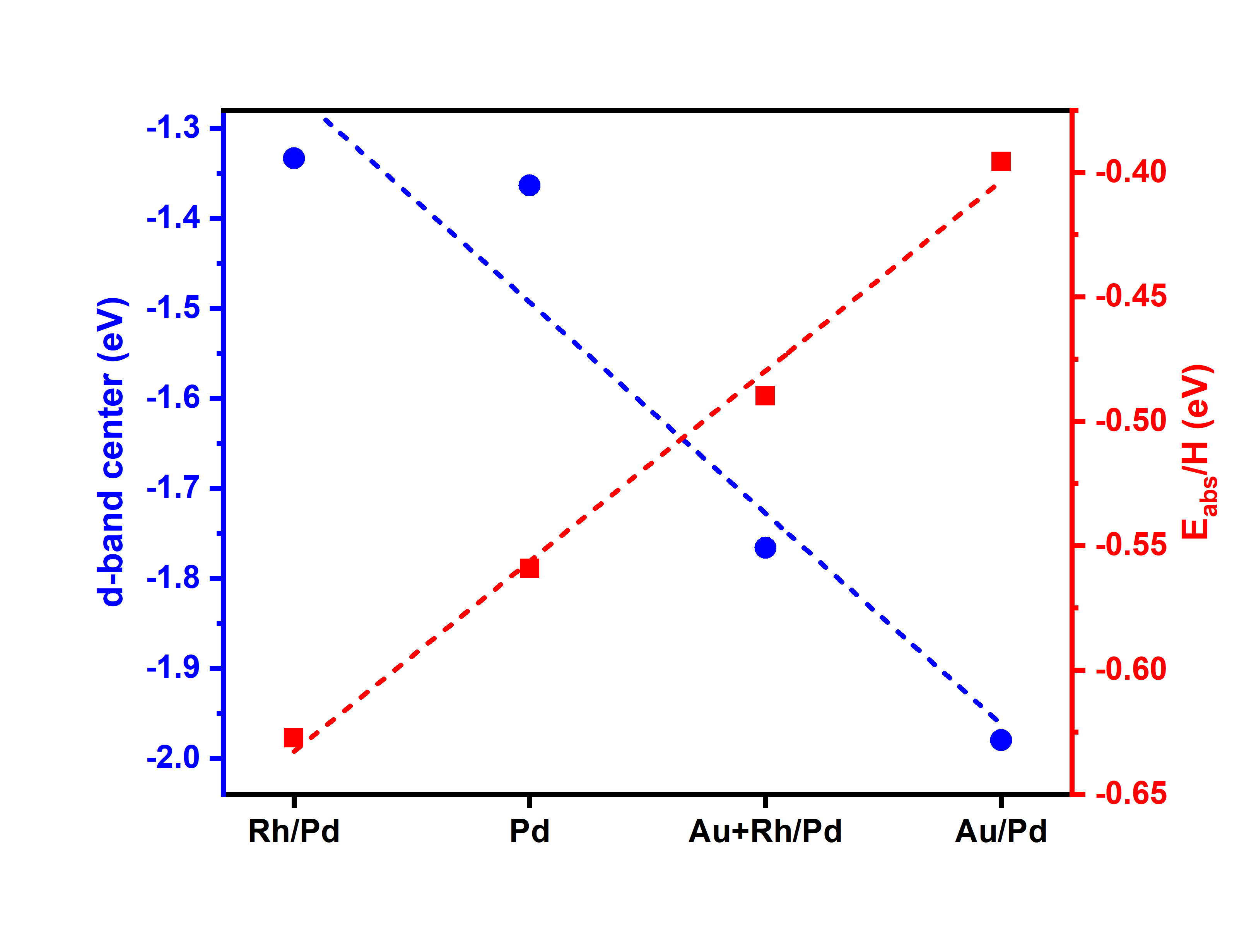}
	\caption{Correlation between E\textsubscript{ads} of a single hydrogen molecule (red dots) and the d-band centers (blue dots) on different Pd surfaces.}
	\label{fig:2}
\end{figure}

\begin{figure}[h!]
	\centering
	\includegraphics[width=0.75\linewidth]{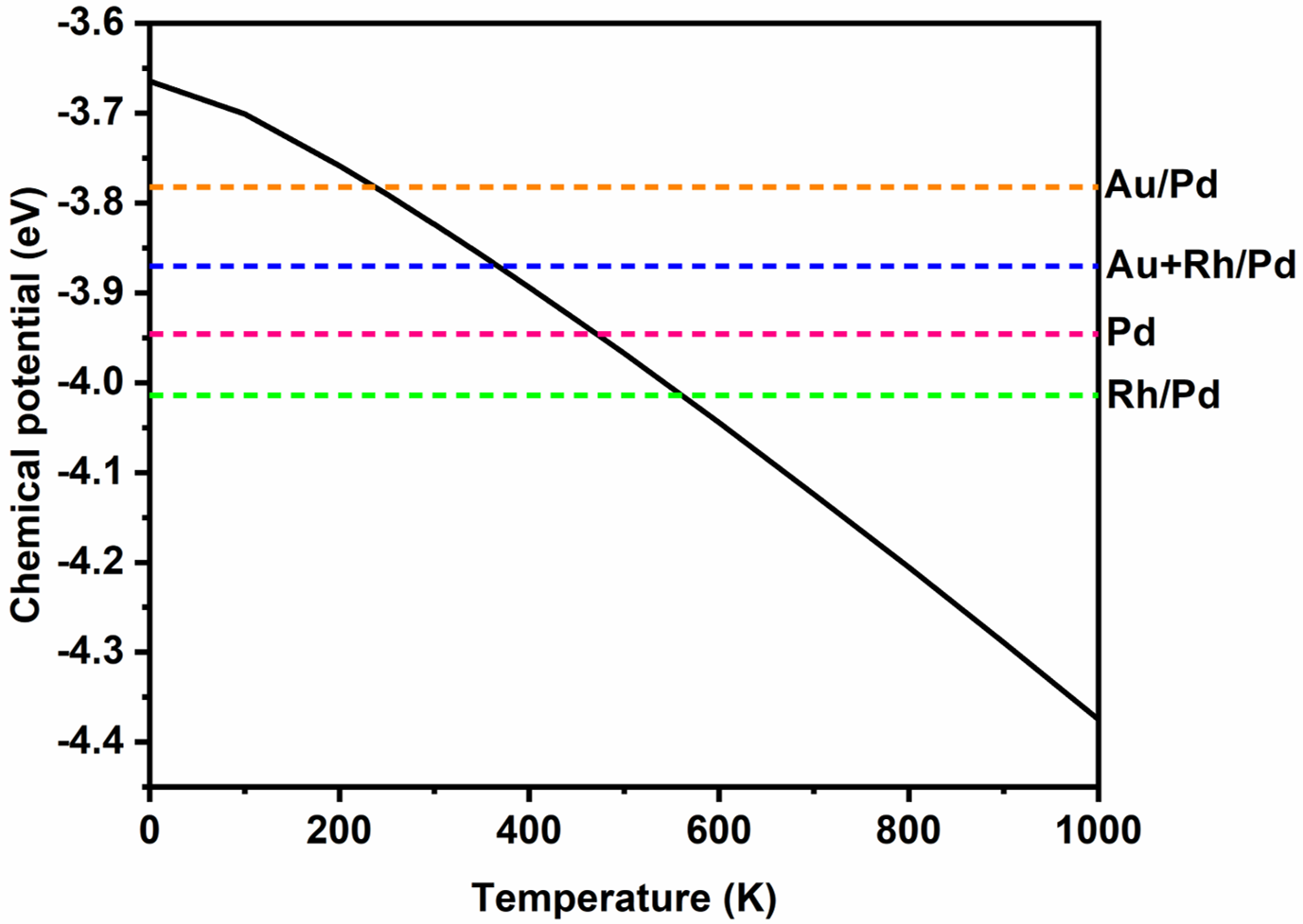}
	\caption{The temperature dependence of the chemical potential of hydrogen (solid black line). Dashed horizontal lines marks the value of $\mu^C_H$ different Pd surface.}
	\label{fig:3}
\end{figure}

In the Fig.\ref{fig:3} we show the variation of the chemical potential of hydrogen atoms with temperature at 1 bar pressure. The value of $\mu^c_H$ for different Pd surfaces is plotted in dashed lines. The intersection between the lines and the hydrogen chemical potential curve indicates the corresponding critical temperature of hydrogen release from the particular surface at 1 bar pressure \cite{Williamson2004}. The temperatures below this intersection will favour hydrogen chemisorption while temperatures above the intersection will favour the desorption (release) of hydrogen. From the figure, it can be seen that for the clean Pd (110) surface, this release temperature is at 477 K. Whereas, for the most studied Au/Pd surface release hydrogen at 228 K at standard atmospheric pressure \cite{Namba2018, EkborgTanner2021}. The hydrogen release temperature of both surfaces is either too high or too low for the ideal operating temperature of a hydrogen fuel cell, which is 340 to 360 K \cite{Eberle2012}. Therefore, both the surfaces are not suitable for hydrogen storage applications. The Rh alloyed Pd surface has the highest hydrogen release temperature (559 K). But when both  Au and Rh are used to surface alloying, it can be seen that an almost ideal 365 K release temperature is obtained. \\

\begin{figure}[h!]
	\centering
	\includegraphics[width=1.0\linewidth]{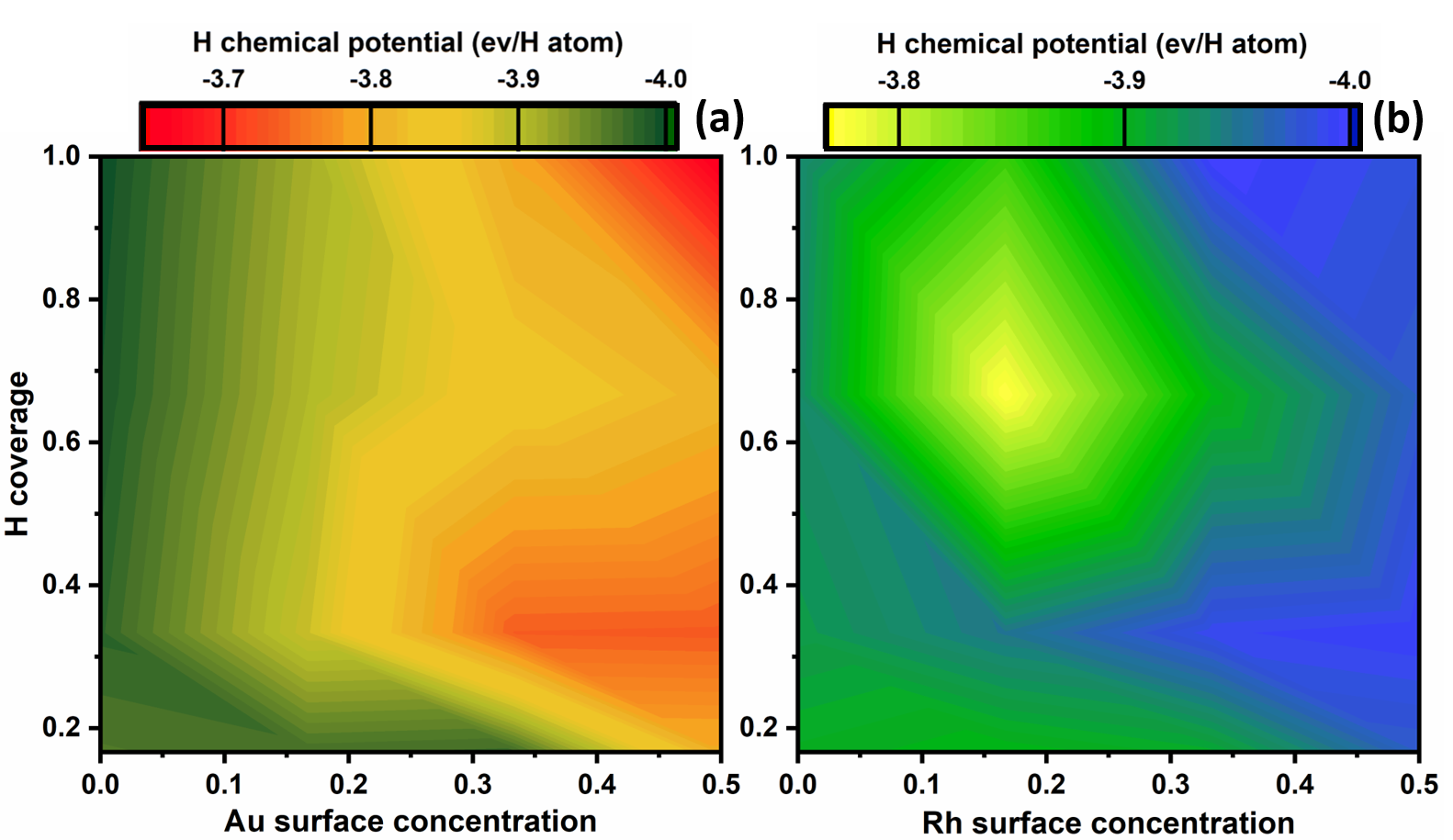}
	\caption{Hydrogen chemical potential as function of H coverage and Au (a) and Rh (b) concentration on Pd (110) surface.}
	\label{fig:4}
\end{figure}

To emphasis it further, we plotted the H chemical potential as a function of Au and Rh surface concentration and the H coverage, as shown in Fig.\ref{fig:4} , to better understand the origin of such perfect behaviour. In this figure, up to 0.5 ML of the surface Pd atoms are replaced by either Au or Rh atoms.
It can be seen that when Pd is alloyed with Au, the critical hydrogen chemical potential changes linearly with Au concentration at fixed H coverage (Fig.\ref{fig:4}(a)). With increasing Au concentration on the Pd surface, the surface loses its affinity to H atoms. However for a fixed Au concentration, the surface chemical affinity to H almost remains same with increasing H atoms on the surface. On the other hand, when Pd is alloyed with Rh, the hydrogen chemical potential changes periodically with Rh concentration at a given H coverage (Fig.3 (b)). But unlike the case of Au alloyed surface, the affinity to hydrogen is stronger at higher Rh concentration. Therefore, in terms of surface H affinity with concentration, alloying with Au and Rh produce exactly opposite effect. This behaviour is actually reflected in terms of the release temperature, which was discussed above.  Therefore for an optimal performance one has to consider a mixed alloy containing both Rh and Au.\\

\begin{figure}[h!]
	\centering
	\includegraphics[width=0.70\linewidth]{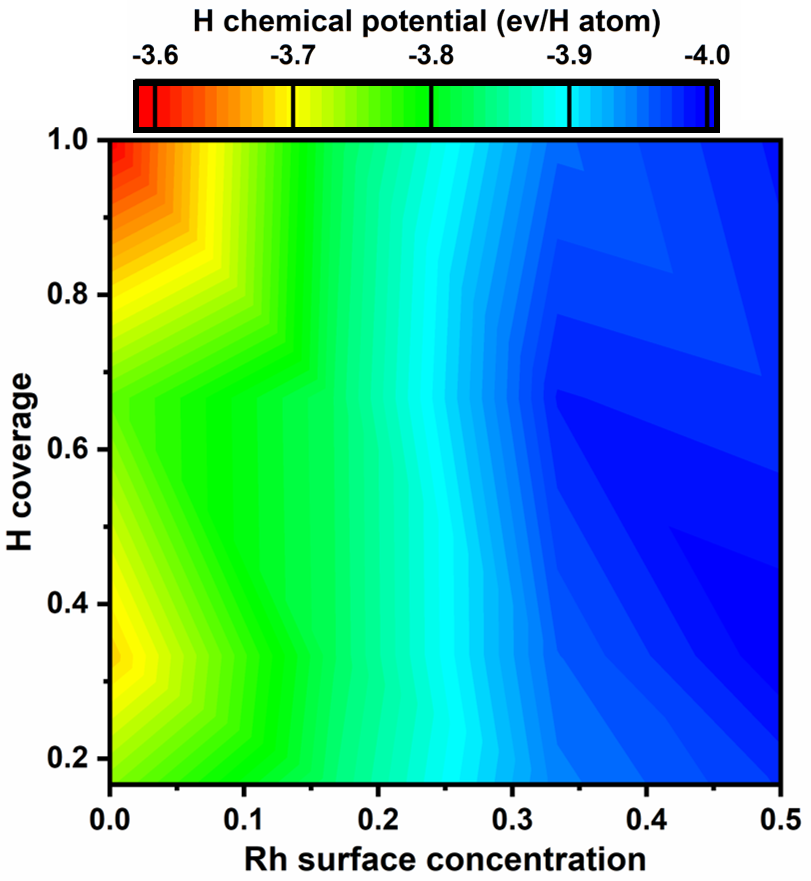}
	\caption{Hydrogen chemical potential as function of H coverage and Au and Rh on Pd (110) surface.}
	\label{fig:5}
\end{figure}

Now, to  illustrate this \textit{optimal concentration}, we depict the scenario where Au and Rh are added to the surface of Pd (110) in the Fig.\ref{fig:5}. Here, 0.5 ML of Au is equal to zero on the x-axis. Rh is substituted for Au along the x-axis until there is 0.5 ML of Rh. It can be shown from a comparison of Fig.\ref{fig:3} and Fig.\ref{fig:5} that the ideal concentration corresponds 0.2 ML Rh and 0.3 ML Au.

\begin{figure}[h!]
	\centering
	\includegraphics[width=0.8\linewidth]{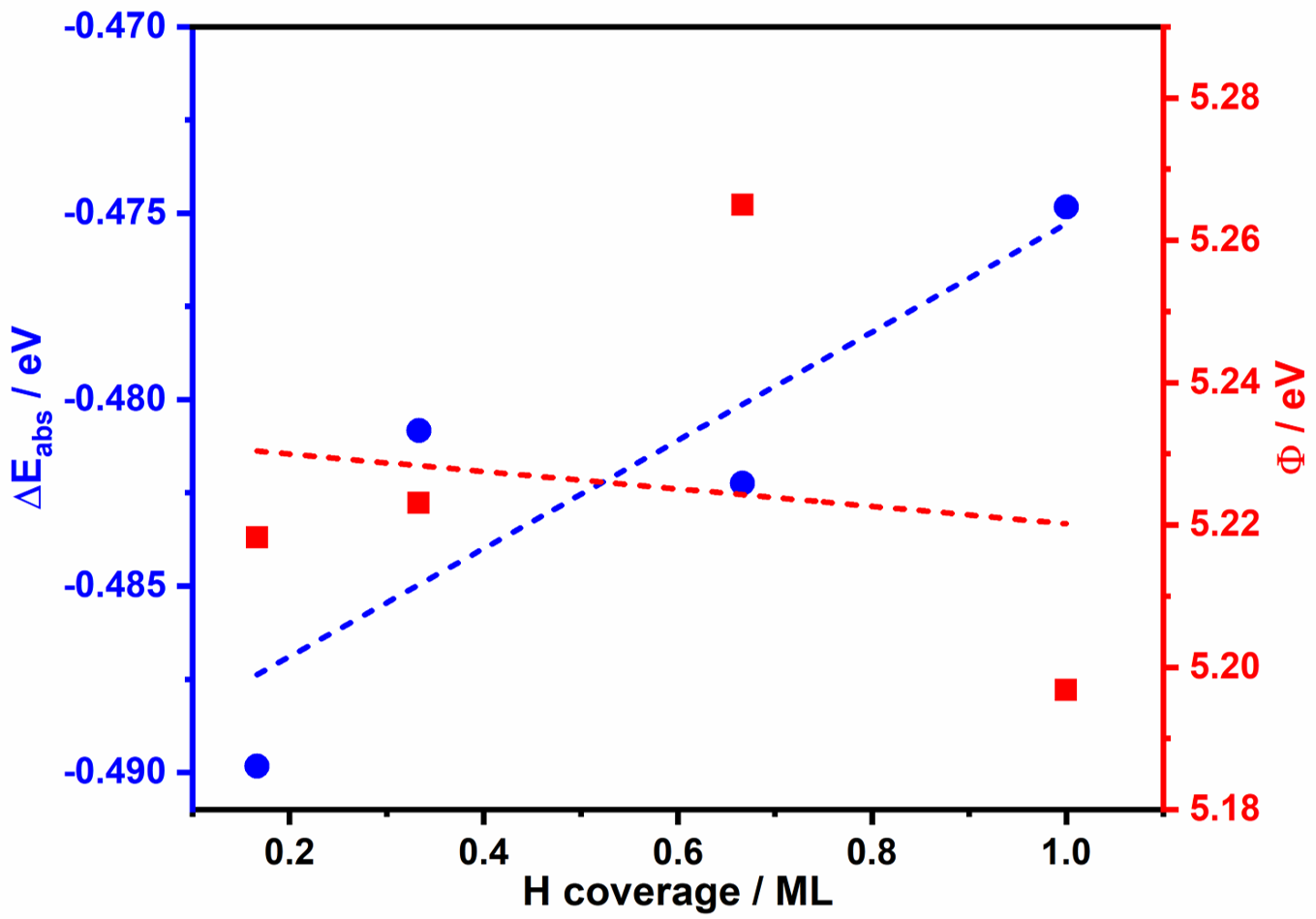}
	\caption{Correlation between hydrogen loading and E\textsubscript{ads} of a single hydrogen molecule (blue) and hydrogen loading and  work functions ($\phi$) (red) on Au+Rh/Pd surfaces.}
	\label{fig:6}
\end{figure}

To study the effect of hydrogen loading, we choose the Au+Rh/Pd surface, as it has almost the ideal hydrogen release temperature. For this purpose, all possible adsorption sites in the top monolayer are gradually filled with a hydrogen atom. A clean and alloyed Pd surface has six adsorption sites. Therefore, the loading of 1, 2, 4, and 6 hydrogen atoms on the Pd surfaces is studied and the results are shown in Fig.\ref{fig:6}.\\

With increase in hydrogen coverage leads to a decrease in the adsorption energy per atom, which is consistent with the results found by previous work \cite{POSADAPEREZ2017}. This effect is not unexpected, and insight comes from the analysis of the work function ($\phi$), which decreases with hydrogen coverage on the Au+Rh/Pd alloy surface as more electrons are filled the conduction band (Fig.\ref{fig:6}) \cite{Silveri2019}.

\section*{Conclusions}

In summary, first-principles calculation has been performed to design a new Pd based alloy, keeping in mind the requirements of hydrogen fuel cell use in ambient atmospheric conditions. A new Au+Rh/Pd alloy structure has been proposed which has 0.49 eV adsorption energy and a 365 K release temperature. Both adsorption energy and release temperature are almost in the range of ideal conditions. Previous studies show the addition of 0.5 ML Au on the Pd (110) surface can enhance the hydrogen absorption capacity, which is important for hydrogen storage. This study shows the addition of Rh with Au can also enhance the adsorption and release mechanism of the hydrogen fuel cell. This finding will lead to improving and controlling the hydrogen storage on Pd alloys for fuel cell applications.

\bibliography{mybibfile}

\end{document}